\documentclass[twocolumn]{revtex4}
\usepackage{amssymb}
\usepackage{latexsym}
\usepackage{epsfig}
\usepackage{graphicx}
\usepackage{amsmath}
\usepackage{lipsum}
\usepackage{xcolor}
\begin{document}
\title{Minimal length implications on the Hartree-Fock theory}
\author{M. Mohammadi Sabet$^1$\footnote{m.mohamadisabet@ilam.ac.ir}, H. Moradpour$^2$\footnote{hn.moradpour@maragheh.ac.ir}, M. Bahadoran$^3$\footnote{bahadoran@sutech.ac.ir}, A. H. Ziaie$^2$\footnote{ah.ziaie@maragheh.ac.ir}}
\address{$^1$ Basic Science Faculty, Physics department, Ilam University, P. O. Box, 69315-516, Ilam, Iran\\
$^2$ Research Institute for Astronomy and Astrophysics of Maragha
(RIAAM), University of Maragheh, P.O. Box 55136-553, Maragheh,
Iran\\
$^3$ Department of Physics, Shiraz University of Technology,
31371555, Shiraz, Fars, Iran}

\begin{abstract}
Hartree-Fock approximation suffers from two shortcomings including
$i$) the divergence of the electron Fermi velocity, and $ii$) the
existence of bandwidth which is not confirmed experimentally.
Here, we study the effects of the minimal length on the ground
state energy of the electron gas in the Hartree-Fock
approximation. Our results indicate that, mathematically, the
correction of minimal length to the phase space, which plays a
vital, and predominant role below the Fermi surface, eliminates
the weaknesses of the Hartree-Fock approximation. On the other
hand, the effect of the Hamiltonian correction, which has the same
form as the relativistic correction of electrons in solids,
becomes dominant at energy levels above the Fermi surface.
Physically, it is concluded that electrons in metals may be
employed to test the quantum gravity scenario, if the value of its
parameter ($\beta$) lies within the range of $2$ to $10$,
depending on the used metal. Indeed, the latter addresses an upper
bound on $\beta$ parameter which is comparable with previous works
meaning that these types of systems may be employed as a benchmark
to examine quantum gravity scenarios. To overcome the Fermi
velocity divergence in the Hartree-Fock method, the screening
potential is used based on the Lindhard theory. In the context of
this theory, we also find that considering the generalized
Heisenberg uncertainly leads to some additional oscillating terms
in the Friedel oscillations.
\end{abstract}

\maketitle


\section{Introduction}

The quantum features of gravity, and indeed quantum gravity, are
amongst the most intriguing challenges of contemporary physics. It
claims that canonical coordinates $x_i$ and $p_i$, satisfying
$[x_i,p_j]=i\hbar\delta_{ij}$, do not necessarily preserve their
ordinary full meanings at the Planck scales, and they should be
replaced by generalized coordinates $X_i$ and $P_j$ for which
$[X_i,P_j]\neq i\hbar\delta_{ij}$~\cite{001,002} see
also~\cite{Hossenfelder} for recent review. In this manner,
Heisenberg uncertainty principle (HUP) is generalized as

\begin{eqnarray}\label{GUP}
(\Delta X)(\Delta P)\geq\frac{\hbar}{2}(1+\beta (\Delta
P)^{2}+...),
\end{eqnarray}
where $\beta$ is called the GUP (generalized uncertainty
principle) parameter, and a non-zero minimum (comparable to the
Planck length) is obtained for $\Delta x$ as $\hbar\sqrt{\beta}$.
Bearing quantum mechanics in mind, it is obvious that replacing
HUP with GUP affects everything such as classical systems and
orbit problem \cite{01,02,03,04,05}, Schr\"{o}dinger equation
\cite{002,1,2,3}, cosmological as well as astrophysical
scenarios~\cite{GUPCOS}, high energy physics \cite{4,5,6} and
optics \cite{optic}. In this regard, Eq.~(\ref{GUP}) implies the
relations $X_i=x_i$, and $P_i=p_i(1+\beta p^2)$ (up to the first
order of $\beta$) which are valid between different mentioned
coordinates that come from the commutation relation
$[X_i,P_j]=i\hbar(1+\beta P^2)\delta_{ij}$ \cite{002}.
\par
Addressed works and other similar attempts such as \cite{me,me2}
have at least two achievements $i$) they study the effects of
existence of a non-zero minimum length on different physical
scales, and $ii$) they give us an estimation on the power and
usefulness of different setups for testing $\beta$ parameter and
quantum features of gravity in different experiments. Despite all
these efforts, there is still a great difference between
theoretical predictions and hypothetical experimental constraints
on the value of $\beta$ parameter~\cite{me2}. Of course, it seems
that the light twisted by rotating black holes may help us find
considerable upper bounds on $\beta$ values compared with those
proposed by quantum mechanics~\cite{fab,me2}. The latter means
that it is still a problem to find a quantum mechanical system
which assists us to verify GUP within Earth-based labs.
\par
Moreover, it seems that there is a deep connection between quantum
features of gravity and generalized statistics~\cite{homa},
where the latter has been
investigated in condensed matter systems such as electrons in
metals~\cite{gen}. On the other
hand, the application of GUP in ideal gas, as the simplest
many-body system, has been investigated in detail
in~\cite{idealgas}, however, the
GUP effects have not been studied for a non-ideal gas. The above
arguments motivate us to study the effects of GUP (the existence
of a non-zero minimum uncertainty in position) on the behavior of
electrons within metals. This
helps us go beyond ideal gas model and makes a new insight to find
the minimal length effects for a more realistic model within the
interacting many body systems. It is therefore expectable that one
could figure out more details of quantum gravity effects in
condensed matter physics.
\par
Hartree-Fock method \cite{HF1,HF2} is one of the most important
theories in physics, especially in metals, in which the $N$-body
wave function is often approximated by the Slater determinant of
$N$ single-particle wave functions. This self consistent method
can be considered as a single particle method and the
inter-particle interaction is studied as a mean field potential.
In spite of good and  accurate results and the applicability  of
this method in metals, there are two important deficiencies:
divergence of velocity at the Fermi surface and prediction of some
values of bandwidth that are not confirmed by
experiments~\cite{martin,modHF2}. However, motivated by the
remarkable achievements of Hartree-Fock approach which includes
interaction as well, our aim in the present work is to investigate
the influences of GUP on this method with the hope of finding a
 test bed for GUP within similar experimental systems. This
could possibly shed some light on the footprints of GUP in
Laboratory studies. The current paper is then organized as
follows: In sections \ref{GUPsec} and \ref{HF} a brief discussion
on GUP formalism along with Hartree-Fock method will be presented,
respectively. The effects of minimal length on HF are investigated
in Sec. \ref{min_length}, and subsequently, the Lindhard screening
theory will be examined in the GUP formalism in the section
\ref{GUP_screening}. Finally, a summary and conclusion will be
presented.
%%%%%%%%%%%%%%%%%%%%%%%%%%%%%%%%%%%%%%%%%%%%%%%%%%%%%%%%%%%%%%%%%%%%%%%%%%%%%%%%%%%%%%%%
%
%
\section{Generalized Uncertainty Principle}\label{GUPsec}
As mentioned previously, GUP relation Eq.~\eqref{GUP} takes the following form,
\begin{equation}\label{GUP1}
[X_i,P_j]=i\hbar(1+\beta P^2)\delta_{ij}.
\end{equation}
Defining the general transformation as
$$(x_i,p_i) \to \left( X(x_i,p_i),P(x_i,p_i)\right),$$
and after doing some algebra, one can get the $N$-dimensional
density of state as \cite{shababi2},
\begin{equation}\label{DOS}
a(\varepsilon)d\varepsilon=\frac{1}{N}\frac{d^NXd^NP}{[\hbar(1+\beta P^2)]^N}.
\end{equation}
Therefore, GUP changes the structure of phase space with a greater
volume element.

On the other hand, following \cite{pedram} and references therein,
we can consider
\begin{equation}\label{momentum_GUP}
P=\frac{\tan(\sqrt\beta p)}{\sqrt\beta},
\end{equation}
as a consequence of GUP relation. Hence, the Schr\"{o}dinger equation can be written as,
\begin{eqnarray}\label{schro_GUP}
H&=&H_0\\\nonumber
&+&\sum_{n=3}^{\infty}\frac{(-1)^{n-1}2^{2n}(2^{2n}-1)(2n-1)B_{2n}}{2m(2n)!}\beta^{n-2}p^{2(n-1)},
\end{eqnarray}
where $H_0=p^2/2m+V(x)$ and $B_n$ is the $n$th Bernoulli number.
It can be shown that in the presence of $n\ge3$ terms, there is a
positive shift within the energy spectrum of particles. It is
worth mentioning that a similar procedure can be applied to the
Dirac equation and more details can be found
in~\cite{nozarikarami}.
\section{Hartree Fock Approximation}\label{HF}

The Schr\"{o}dinger equation governing the behavior of electrons
in metals is given as follows
\begin{equation}\label{Schro_eq}
 H_0\Psi ({{\bf{r}}_1},....,{{\bf{r}}_N}) = E\Psi
({{\bf{r}}_1},....,{{\bf{r}}_N}),
\end{equation}
with
\begin{equation}\label{Hamiltonian1}
H_0 = \sum\limits_{i = 1}^N {\left( { - \frac{{{\hbar
^2}}}{{2m}}\nabla _i^2 - Z\sum\limits_R
{\frac{{{e^2}}}{{|{{\bf{r}}_i} - {\bf{R}}|}}} } \right)}  +
\frac{1}{2}\sum\limits_{i \ne j}^N {\frac{{{e^2}}}{{|{{\bf{r}}_i}
- {{\bf{r}}_j}|}}},
\end{equation}
where the first and the second terms are kinetic energy and ionic
interaction potential energy, respectively and the last term deals
with the electron-electron interaction. Hartree-Fock
method~\cite{HF1,HF2} is one of the most important approaches to
obtain the effects of electron-electron interactions within
metals. In this approximation all correlations are neglected
except that of the Pauli exclusion principle and  the $3$rd term
in the Hamiltonian is converted to a single particle form
considering the effect of other electrons as a smooth negative
charge distribution. Consequently, utilizing the Slater
determinant of single-particle wave functions, the Hamiltonian
will be rewritten as a set of $N$ one-body problems, one of them
for each one-electron level. More details can be found in
\cite{HF2}. Considering then the plane wave as the single-electron
wave function, the energy takes the following form
\begin{equation}\label{en_1}
\varepsilon ({\bf{k}}) = \frac{{{\hbar ^2}{k^2}}}{{2m}} - \frac{1}{V}\sum\limits_{k'} {\frac{{4\pi {e^2}}}{{|{\bf{k}} - {\bf{k'}}|}}},
\end{equation}
where $\frac{{4\pi {e^2}}}{{|{\bf{k}} - {\bf{k'}}|}}$ is the Fourier transform of the exchange energy. Now using
\begin{equation}\label{sum_rul}
\sum\limits_{\bf{k}} { \to \frac{V}{{{{(2\pi )}^3}}}} \int
{d{\bf{k}}},
\end{equation}
the expression for energy can be rewritten as
\begin{equation}\label{one-body_HF}
\varepsilon ({\bf{k}}) = \frac{{{\hbar ^2}{k^2}}}{{2m}} - \frac{{2{e^2}}}{\pi }{k_{\rm F}}{F_0}\left(\frac{k}{{{k_{\rm F}}}}\right),
\end{equation}
where
\begin{equation}\label{F_0}
F_0(x)=\frac{1}{2}+\frac{1-x^2}{4x}\ln\Big|\frac{1+x}{1-x}\Big|.
\end{equation}
Here, $x$ is defined as the ratio $\frac{k}{k_{\rm F}}$. The
energy per particle can be calculated through the summation of Eq.
\eqref{one-body_HF} over ${\bf k}\le{\bf k_F}$ along with using
Eq. \eqref{sum_rul}. We therefore get
\begin{eqnarray}\label{E_N}
\frac{E}{N} &&= \frac{{{e^2}}}{{2{a_0}}}\left[ {\frac{3}{5}{{({k_{\rm F}}{a_0})}^2} - \frac{3}{{2\pi }}({k_{\rm F}}{a_0})} \right]\\\nonumber
&&= \left[ {\frac{{2.21}}{{{{({r_s}/{a_0})}^2}}} - \frac{{0.916}}{{({r_s}/{a_0})}}} \right]Ry,
\end{eqnarray}
where $a_0$ is the Bohhr radius and $r_s$ is defined as
$r_s=\left(\frac{1}{4\pi n}\right)^{1/3}$ where $n=\frac{k_{\rm
F}^3}{3\pi^2}$ and $k_{\rm
F}=\left(\frac{3.63}{r_s/a_0}\right)\AA^{-1}$.

\section{Minimal Length Effects on Hartree-Fock Energy }\label{min_length}
In order to calculate the effects of minimal length on the energy
of  the electron liquid in metals, we use, firstly, Eq.
\eqref{schro_GUP} upto the first term in the summation. Therefore,
we have, instead of Eq. \eqref{en_1} the following equation for
the Fourier transform of single particle energy,
\begin{equation}\label{en_1_R}
\varepsilon ({\bf{k}}) = \frac{{{\hbar ^2}{k^2}}}{{2m}}+\beta\frac{ \hbar^4 k^4}{3m} - \frac{1}{V}\sum\limits_{k'} {\frac{{4\pi {e^2}}}{{|{\bf{k}} - {\bf{k'}}|}}},
\end{equation}%
where $\textbf{k}$ and $\textbf{k}'$ are ordinary momenta. Therefore, after integrating over all $k$ states, we obtain%
\begin{equation}\label{En2_4}
\frac{E}{N} = \frac{{2.21}}{{{{({r_s}/{a_0})}^2}}} - \frac{{0.916}}{{({r_s}/{a_0})}}+  \beta \frac{{20.03}}{{{{({r_s}/{a_0})}^7}}} \ Ry,
\end{equation}
where we have used Eq. \eqref{sum_rul}. As it is clear from Eq.
\eqref{En2_4}, by considering the perturbed Hamiltonian, Eq.
\eqref{schro_GUP} and $d^3xd^3p$ as the phase space volume (where
$x$ and $p$ are canonical coordinates), the GUP correction has no
contribution in the interaction and exchange energy and the
minimal length affects only the kinetic energy as an $x^4$ term.
This term is a high density term (small $r_s$)  due to the power
of $r_s/a_0$.

\begin{figure}[!h]
    \includegraphics[scale=.7]{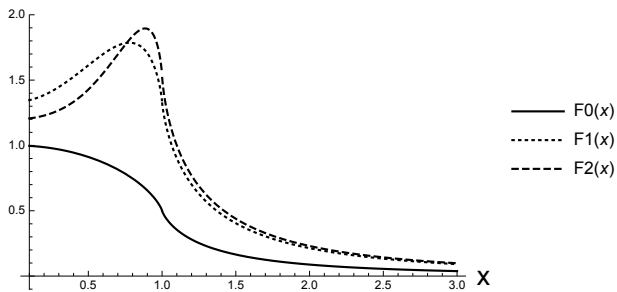}%
    \caption{Factor function for different orders of $\beta$.}%
    \label{correction}
\end{figure}

In order to get more details and for a complete investigation, one
must use Eq. \eqref{schro_GUP} together with $d^3Xd^3P$ as the
generalized phase space volume. Therefore, up to the first order
of $\beta$ parameter, The Hamiltonian gets the following form,

    \begin{equation}\label{hamiltonian_gup_1}
\mathcal{\hat H}=H_0+\sum_{i}\frac{\beta}{3m}p_i^4+\mathcal{O}(\beta^2)+\dots,
\end{equation}
To proceed one must use \cite{MAZ,pedram}
\begin{equation}\label{GUP_sum_rul}
\sum\limits_{\bf{P}} { \cdots  \to \frac{V}{{{{(2\pi )}^3}}}} \int { \cdots {{(1 + \beta{P^2})}^{ - 3}}d{\bf{P}}}.
\end{equation}
Considering the first order of $\beta $ parameter in Eq. \eqref{momentum_GUP}, we have,
\begin{equation}\label{mmentum_gup2_1}
P=p\left(1+\frac{1}{3}\beta p^2\right).
\end{equation}
Therefore, Eq. \eqref{en_1} must be replaced by
\begin{eqnarray}\label{en_1_gup_1}\nonumber
\varepsilon ({\bf{k}}) &=& \frac{{{\hbar ^2}{k^2}}}{{2m}} +\frac{\beta}{3m}\hbar^4k^4- \frac{{4\pi {\hbar ^2}}}{{{{(2\pi \hbar )}^3}}}\int {\frac{{{{(1 + \beta \,{{P'}^2})}^{ - 3}}}}{{|{\bf{p}} - {\bf{p'}}{|^2}}}d{\bf{P'}}} \\
%&=& \frac{{{\hbar ^2}{k^2}}}{{2m}} - \frac{{{e^2}}}{{\pi \hbar p}}\int {\frac{{p'\ln\left| {\frac{{p' + p}}{{p' - p}}} \right|}}{{{{(1 + \beta \,{{p'}^2})}^3}}}dp'}.
\end{eqnarray}
Expanding $(1+\beta P'^2)^{-3}$ together with Eq.
\eqref{mmentum_gup2_1} and $dP=dp+\beta p^2dp$ as well as
$P^2=p^2+2/3\beta p^4$ (up to the first order of $\beta$), we get
\begin{equation}\label{sum_rul_1}
\int \dots \frac{d\textbf{P}}{(1+\beta P^2)^3}=\int \dots (1-4/3\beta p^2)d\textbf{p}.
\end{equation}
After some algebra, we have
\begin{eqnarray}\label{en_1_GUP2_1}
\frac{{\varepsilon ({\bf{\rm k}})}}{{\varepsilon_{\rm F}^0}} &=& {x^2}+\beta (4.843(r_s/a_0)^{-2})x^4\\\nonumber
&-& 0.663({r_s}/{a_0}){F_0}(x)\\ \nonumber
&-&\beta\ 2.42 (r_s/a_0)^{-1}F_1(x) \\\nonumber
&+&C_2{\beta ^2}{({r_s}/{a_0})^{ - 3}}{F_2}(x)+\mathcal{O}(\beta^3)+\dots,
\end{eqnarray}
where $C_2$ is a real constant and $\varepsilon_{\rm F}^0$ is the Fermi energy of the ideal electron gas. One can write Eq. \eqref{en_1_GUP2_1}, for simplicity, as follows,
\begin{equation}\label{en_1_gup3}
\tilde \varepsilon ({\bf{\rm k}})=\tilde \varepsilon^{\rm HF} ({\bf{\rm k}})+\tilde \varepsilon^{\rm GUP} ({\bf{\rm k}}),
\end{equation}
where $\tilde \varepsilon ({\bf{\rm k}})=\frac{{\varepsilon
({\bf{\rm k}})}}{{\varepsilon _F^0}}$, $\tilde \varepsilon^{\rm
HF} ({\bf{\rm k}})$ is that of Eq. \eqref{one-body_HF} scaled by
$\varepsilon_F^0$ and
\begin{equation}\label{en_1_GUP3}
\tilde \varepsilon^{\rm GUP} ({\bf{k}})=
\beta\left(-{\rm{2}}{\rm{.42}} {({r_s}/{a_0})^{ -
1}}{F_1}(x)+(4.843(r_s/a_0)^{-2})x^4\right),
\end{equation}
is the GUP corrections on the exchange energy. The correction
factors $F_i(x)$ ($i$ refers to the order of correction) are given
by
\begin{eqnarray}\label{F1F2F3}\nonumber
&&{F_1}(x) = \frac{1}{3} + {x^2} + \frac{{1 -
{x^4}}}{{2x}}\ln\left| {\frac{{1 + x}}{{1 - x}}}
\right|\\\nonumber &&{F_2}(x) = \frac{1}{5} + \frac{{{x^2}}}{3} +
{x^4} + \left(\frac{{1 - {x^6}}}{{2x}}\right)\ln\left| {\frac{{1 + x}}{{1 -
x}}} \right|\\\nonumber &&{F_3}(x) = \frac{1}{{45}} + {x^{ - 1}} +
\frac{{{x^2}}}{7} + \frac{{{x^4}}}{5} \\\nonumber
&&+ \left(\frac{{1 - {x^{10}}}}{{2x}}\right)\ln\left| {\frac{{1 + x}}{{1 - x}}} \right|\\
&& \dots~.
\end{eqnarray}
These factors are plotted in Fig. \eqref{correction}. It is clear
from this figure that the value of correction factors assumes a
significant difference with respect to the zeroth order of $\beta$
parameter. Moreover, at higher values of $x$, the difference
decreases and tends to a minimum value so that all these factors
are nearly in the same order.

In order to investigate the contribution of these corrections to
the total energy of the $N$-particle system, we must add up these
corrections to all values of wave vectors which are below the
Fermi wave vector i.e., $\bold{k}\le \bold{k}_F$. Considering GUP
summation rule i.e., Eq. \eqref{sum_rul_1} together with making
sum of Eq. \eqref{en_1_GUP2_1} and after some algebra, the total
energy per particle, up to the first order of $\beta$ parameter,
can be written as
%%%================================
\begin{equation}\label{gup_tot_en}
\begin{array}{l}
\frac{E}{N} = \frac{{2.21}}{{{{({r_s}/{a_0})}^2}}} - \frac{{0.916}}{{({r_s}/{a_0})}}\\
\,\,\,\,\,\, + \beta \left( {\frac{{1.0829}}{{{{({r_s}/{a_0})}^3}}} + \frac{{3.92}}{{{{({r_s}/{a_0})}^4}}} - \frac{{20.04}}{{{{({r_s}/{a_0})}^7}}}} \right)\ Ry,
\end{array}
\end{equation}
whence we get GUP-HF-exchange energy as follows
\begin{equation}\label{gup_exc}
\frac{E_{\rm exc}^{\rm GUP-HF}}{N}=\beta \left( \frac{1.0829}{(r_s/a_0)^3}+ \frac{{3.92}}{{{{({r_s}/{a_0})}^4}}} - \frac{{20.04}}{{{{({r_s}/{a_0})}^7}}}\right).
\end{equation}
%============================

\begin{figure}[!h]
    \includegraphics[scale=.6]{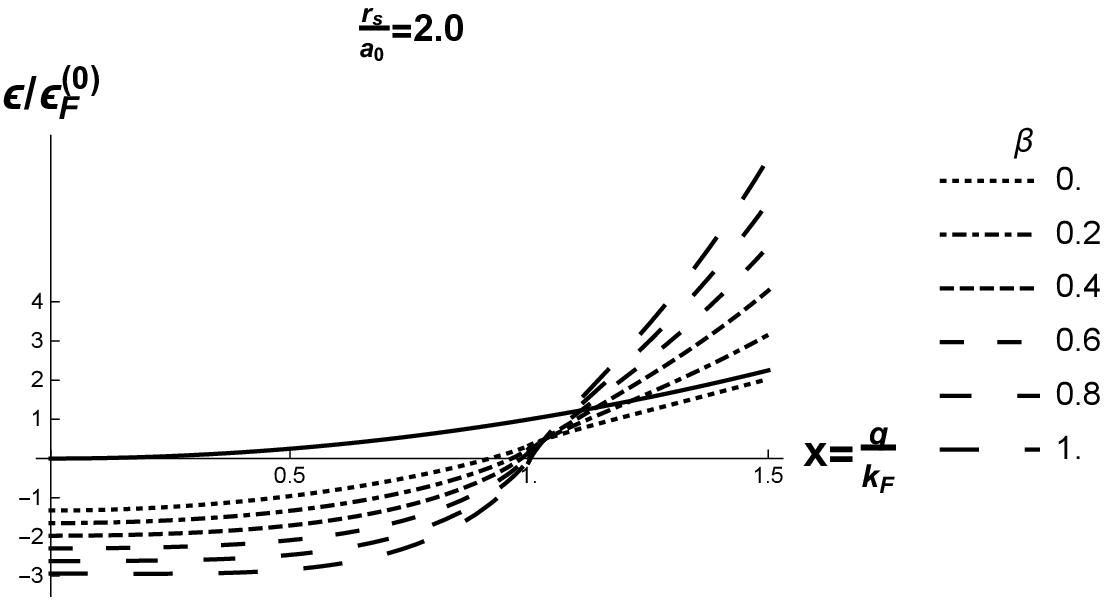}
    \hspace{2cm}
    \includegraphics[scale=.6]{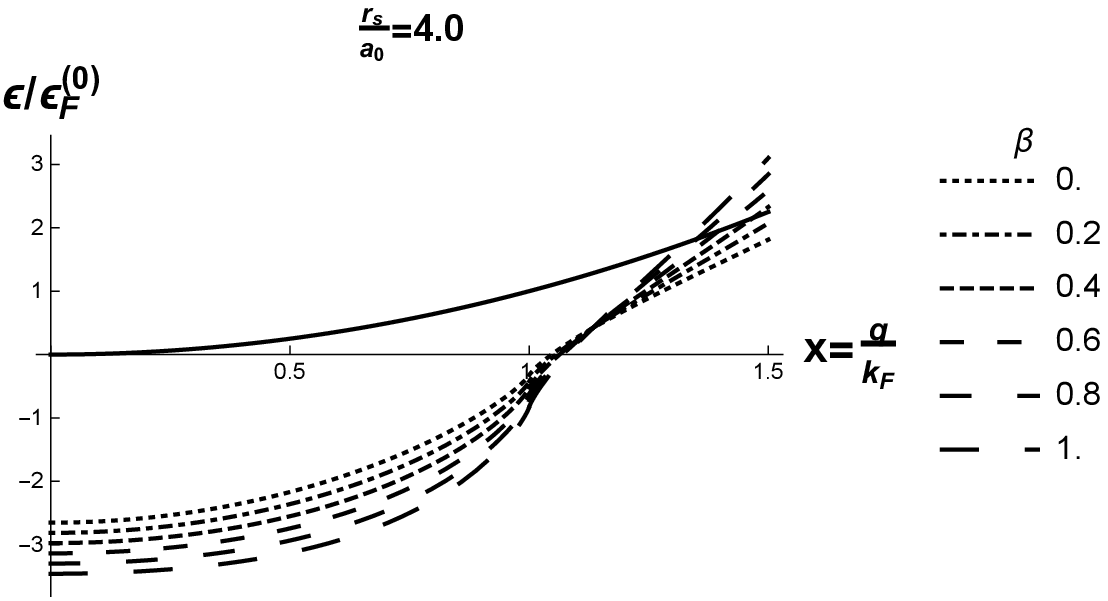}
    \caption{The Hartree-Fock energy with different values of $\beta$ in comparison with free electron gas. The solid curve shows the energy of free electron gas and the $\beta=0.0$ curve indicates the conventional Hartree-Fock method. Other values of $\beta$ present the effect of GUP for $\frac{r_s}{a_0}=2.0$ and $\frac{r_s}{a_0}=4.0$.}
    \label{energy}
\end{figure}

\begin{figure}[!h]
    \includegraphics[scale=.7]{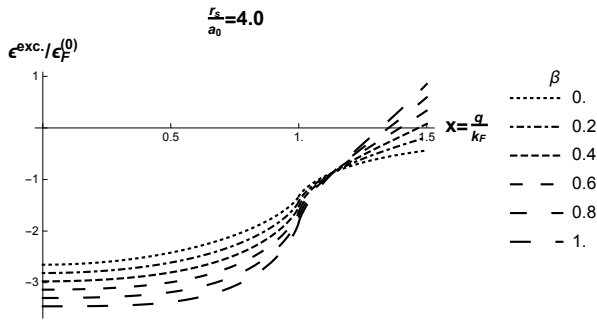}%
    \caption{The exchange energy with different values of $\beta$.}%
    \label{energy_xc}
\end{figure}

Figure~\eqref{energy} shows the behavior of total energy of GUP
corrected Hartree-Fock approximation as a function of $x(=k/k_{\rm
F})$ for different values of $\beta$ and $r_s$ parameters. The
free electron gas energy has been also plotted for comparison. In
this figure, $\beta=0$ refers to Hartree-Fock energy without
considering the minimal length effects. The exchange energy has
been also plotted in Fig.~\eqref{energy_xc} for better
understanding of GUP corrections.  As it is clear from these
figures, the effects of minimal length is more considerable at
higher values of $\beta$ parameter. The minimal length corrections
are less significant as momentum increases, but the GUP effects
can not be ignored at the Fermi surface.

One of the problems in the Hartree-Fock method is the values of
bandwidth for which the predicted values were not confirmed by
experiments, however, as we can see the GUP, in general, increases
the bandwidth value. But, such an increment can be eliminated
considering some values of $\beta$ based on the Fermi velocity.
Another deficiency of the Hartree-Fock method is the divergence of
velocity at the Fermi surface that is inconsistent with
experiments \cite{HF1}. In order to solve this problem a
modification of Hartree-Fock method \cite{modHF1} as well as the
screening theory \cite{HF2} have been proposed.

\begin{figure}[!h]
    \includegraphics[scale=0.7]{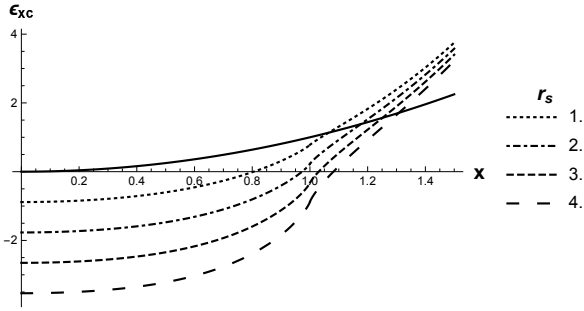}\caption{The GUP-HF energy for different values of $r_s$, The solid curve shows the free electron gas energy. This figure is plotted for $\beta=0.068498\times (r_s/a_0)^2$ }
    \label{Ebond}
\end{figure}

Since this singularity is due to the Logarithmic term in the
$\varepsilon(\bold k)$, it can be eliminated as an another
proposal, namely, choosing appropriate values of $\beta$ parameter
in the GUP version of the Hartree-Fock method. As we know, the
Fermi velocity is defined as,
\begin{equation}\label{Fermi_v1}
\bold{v}_{\rm F}=\frac{1}{\hbar}\nabla_\bold{k}\varepsilon(\bold{k})|_{k=k_{\rm F}}.
\end{equation}
Using the above equation together with Eq. \eqref{en_1_GUP2_1} and
setting the coefficients of Logarithmic terms to zero, after a
little algebra we have
$$\beta=0.06849 \times (r_s/a_0)^2=\mathcal{O}\left((r_s/a_0)^2\right).$$
Hence, we get the following Fermi velocity in the GUP form of the
Hartree-Fock model
$$v_{\rm F}=(4.55\ (r_s/a_0)^{-2}-1.01)\times10^{8}\ cm/s,$$
which is in complete agreement with experimental values
\cite{HF1}. Now this value of $\beta$ parameter is used to
calculate the bandwidth and the exchange energy contribution. The
bandwidth is defined as the energy difference between the energy
at the Fermi level and zero momentum energy
\begin{equation}\label{bandw}
\varepsilon_{\rm B}=\varepsilon_{k_{\rm F}}-\varepsilon(0).
\end{equation}
As a result, one can plot the exchange energy using the above
value for $\beta$ parameter. Figure \eqref{Ebond} represents the
GUP-HF energy for $\beta=0.068498\times (r_s/a_0)^2$. The results
show that considering this value of $\beta$ parameter leads to
decreasing the bandwidth to its HF value.
Besides, this bandwidth can be eliminated using the effects of
minimal length, which is concordant to the experiment. To this
end, the energy at points $x \to 1$ and $x=0$ is required which
can be calculated through Eq. \eqref{en_1_GUP2_1}. A
straightforward calculation then gives
$$\begin{gathered}
{F_0}(0) = 1\,\,\,\,\,\,\,\,\,\,\,\,\,\,{F_0}(x \to 1) = 1/2, \hfill \\
{F_1}(0) = 4/3\,\,\,\,\,\,\,{F_1}(x \to 1) = 4/3, \hfill \\
\end{gathered} $$
Substituting these values back into Eq. \eqref{en_1_GUP2_1} and
assuming the bandwidth energy is equal to zero in Eq.
\eqref{bandw}, we reach
$$\varepsilon_{\rm B}=1.0+0.3315 (r_s/a_0)+4.84 (r_s/a_0)^{-2}\beta ^2=0.$$
This equation can again has a solution in second order of
$(r_s/a_0)$ with an appropriate constant.
%=========================
%=========================
%============================================================================================================
%==================================================================================
\begin{figure}[!h]\label{EperN}
    \includegraphics[scale=.7]{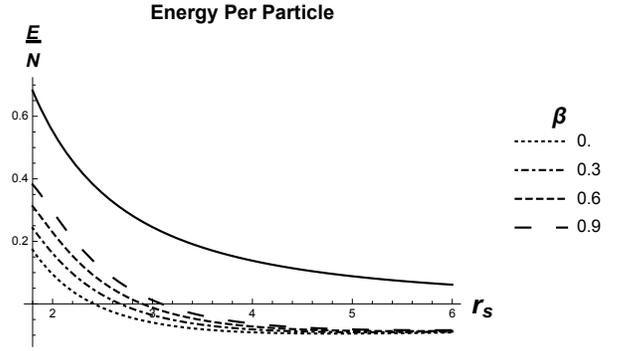}\caption{ Total energy per particle as a function of $r_s$ for different values of $\beta$ parameter. The solid curve shows the energy per particle for free electron gas.}
\end{figure}
%================================================================
%
\section{GUP Effects on Lindhard Theory }\label{GUP_screening}
Another method to eliminate the divergence of Fermi velocity is
the screening theory. In order to consider the screening effect,
we can refer to Thomas-Fermi theory  and the Lindhard theory of
screening \cite{HF1,HF2}. In this manner, we find the charge
density in the presence of the total potential $\phi(r)$ by
solving the one-body Schr\"{o}dinger equation,
\begin{equation}\label{schro1}
\frac{\hbar^2}{2m}\nabla^2\psi _i(r)-e\phi(r)\psi_i(r)=\varepsilon_i\psi_i(r).
\end{equation}
Since the Thomas-Fermi theory of screening has been previously
investigated in the GUP formalism \cite{pedram}, we here evaluate
the minimal length effects on the Lindhard theory of screening.
This theory is based on the fact that the induced density is
linearly proportional to potential $\phi$ . In terms of Fourier
transform, the problem is to find the response function,
$\chi(\textbf{q})$,
\begin{equation}\label{rho_ind1}
\rho^{\rm ind}(\textbf{q})=\chi(\textbf{q})\phi(\textbf{q}),
\end{equation}
where $\rho^{\rm ind}(\textbf{q})$ and $\phi(\textbf{q})$ are the Fourier transform of induced charge density and screening potential, respectively. Calculations show that the dielectric function is related to the response function as well as charge density in the following from
\begin{equation}\label{dielctric1}
\epsilon(\textbf{q})=1-\frac{4\pi}{q^2}\chi(\textbf{q})=1-\frac{4\pi}{q^2}\frac{\rho^{\rm ind}(\textbf{q})}{\phi(\textbf{q})},
\end{equation}
and through a straightforward procedure \cite{HF1}, we get
\begin{equation}\label{LT1}
\chi _{\rm L}=e^2\sum\limits_{\bf{k}} {\frac{{{f^{(0)}}({\bf{k}}) - {f^{(0)}}({\bf{k + q}})}}{{\varepsilon ({\bf{k}}) - \varepsilon ({\bf{k + q}})}}}.
\end{equation}
Therefore, the dielectric function is found as follows
\begin{equation}\label{diel1}
\epsilon({\bf{q}}) = 1 - \frac{{4\pi {e^2}}}{q}\sum\limits_k {\frac{{{f^{(0)}}({\bf{k}}) - {f^{(0)}}({\bf{k + q}})}}{{\varepsilon ({\bf{k}}) - \varepsilon ({\bf{k + q}})}}}.
\end{equation}
After some algebra we get

\begin{equation}\label{LT2}
\chi_{\rm L}(q)=- \frac{{{k_{\rm F}}}}{{{\pi ^2}}}\left[ {\frac{1}{2} + \frac{{1 - 4{x^2}}}{{4x}}\ln \left| {\frac{{1 + x}}{{1 - x}}} \right|} \right],
\end{equation}
where $x=\frac{q}{2k_{\rm F}}$ and therefore the static dielectric constant in Lindhard theory becomes
\begin{equation}\label{diel2}
\epsilon({\bf{q}})=1+\frac{4\pi}{q^2}\frac{{{k_{\rm F}}}}{{{\pi ^2}}}\left[ {\frac{1}{2} + \frac{{4k_{\rm F}^2 - q^2}}{{8k_{\rm F}q}}\ln \left| {\frac{{2k_{\rm F} + q}}{{2k_{\rm F} - q}}} \right|} \right].
\end{equation}
Consequently the screen parameter, $\lambda_{\rm L}$ can be shown by
%
%\begin{equation}\label{Screenpar0}
%\lambda_L^2({\bf{q}})=4\pi\frac{{{k_F}}}{{{\pi ^2}}}\left[ {\frac{1}{2} + \frac{{4k_F^2 - q^2}}{{8k_Fq}}\ln \left| {\frac{{2k_F + q}}{{2k_F - q}}} \right|} \right]
%\end{equation}
%
\begin{eqnarray}\label{Screenpar_length}
\lambda_{\rm L}^2({\bf{q}})&=&4\pi\frac{{{k_{\rm F}}}}{{{\pi ^2}}}\left[ {\frac{1}{2} + \frac{{4k_{\rm F}^2 - q^2}}{{8k_{\rm F}q}}\ln \left| {\frac{{2k_{\rm F} + q}}{{2k_{\rm F} - q}}} \right|}  \right]\\\nonumber
&=&4\pi\frac{{{k_{\rm F}}}}{{{\pi ^2}}}\left[ {\frac{1}{2} + \frac{{1 - x^2}}{{4x}}\ln \left| {\frac{{1+x}}{{1-x}}} \right|} \right].
\end{eqnarray}
These equations admit a singularity at $q=2k_{\rm F} (x=1)$ and
this singularity is a realistic effect due to the summation over
states $\left| \bf{k} \right\rangle $ and $\left| \bf{k}+\bf{q}
\right\rangle$ in Eq. (\ref{LT1})
($f^{(0)}(\bf{k})-f^{(0)}(\bf{k}+\bf{q})$). The main feature of
this singularity is a very interesting effect known as Friedel or
Ruderman-Kittel oscillations. Based upon the Lindhard theory, at
large distances the screened potential $\phi$ of a point charge is
more structured than the simple Thomas-Fermi (Yukawa) screened
potential
\begin{equation}\label{Friedle0}
\phi(r)\sim \frac{1}{r^3}\cos(2k_{\rm F}r).
\end{equation}
In order to get this form of potential one can consider a spherical potential $U(r)$ from which the solution of
Schr\"{o}dinger equation is obtained as follows
\begin{equation}\label{sol_schro}
\psi _{k,l}(r,\theta ) = {A_{k,l}}\frac{1}{r}\sin(kr - l\frac{\pi }{2} + {\delta _l}){P_l}(\cos\theta ),
\end{equation}
and at large distances the change in electron charge density can
be obtained by
\begin{equation}\label{ind_ro}
\delta \rho  = 2\sum\limits_{{\bf{k}} \le {{\bf{k}}_{\rm F}}} {\left[ {{{\left| {\begin{array}{*{20}{l}}
                    {{\psi _k}(r,t)}
                    \end{array}} \right|}^2} - {{\left| {\psi _{_k}^0(r)} \right|}^2}} \right]},
\end{equation}
where $\psi _{_k}^0(r)$ is the plane wave solution of the
unperturbed system. After some boring algebra and using
Eq.~\eqref{sum_rul} we finally get
\begin{equation}\label{friedel1}
\delta \rho  \approx \frac{{ - e}}{{2{\pi ^2}}}\sum\limits_l {\left[ {(2l + 1){{( - 1)}^l}\sin{\delta _l}\frac{{\cos \left( {2{k_F}r + {\delta _l}} \right)}}{{{r^3}}}} \right]} .
\end{equation}
Since, the screen potential is proportional to the change in the
charge density \cite{HF2} we arrive at Eq. \eqref{Friedle0}.

\subsection{GUP Lindhard Screening}
To study the effect of the minimal length on Friedel oscillations
and screen potential, the procedure of calculations is similar to
that of  previously discussed. First of all, to find the GUP form
of $\chi_{\rm L}$, one should use Eqs. (\ref{GUP_sum_rul}) and
(\ref{LT1}), and after a few calculations we obtain
\begin{equation}\label{LT3}
\chi_{\rm L}^{\rm GUP}=\chi_{\rm L}+\beta\ \chi_{\rm L}^1,
\end{equation}
where $\chi_{\rm L}$ is that of Eq. (\ref{LT2}) and  $\chi_{\rm L}^1$ is the first order correction of minimal length with $x=\frac{q}{2k_{\rm F}}$
\begin{equation}\label{LT4_cor}
\chi_{\rm L}^1=-\frac{2k_{\rm F}^4}{3\pi^2q^2}\left(\frac{1}{2}x^3+\frac{1}{6}x+\frac{1}{2}\ln\left|\frac{1+x}{1-x}\right|-\frac{x^4}{32}\ln\left|\frac{1+x}{1-x}\right|\right).
\end{equation}
Again,  for calculating the effect of minimal length on the
Friedel osculations we can evaluate Eq. (\ref{ind_ro}) considering
the GUP summation rule (Eq. (\ref{sum_rul_1})). By doing so, we
have
\begin{equation}\label{GUP_Friedel1}
\delta \rho^{\rm GUP}=\delta \rho+\beta \delta \rho^{(1)},
\end{equation}
where $\delta \rho$ is that of normal screening (Eq.(\ref{friedel1})) and $\delta \rho^{(1)}$ is the first order correction of
minimal length that can be calculated as follows
%$$
\begin{equation}\label{ind_ro_gup}
\begin{gathered}
\delta {\rho ^{(1)}} =  - \frac{4e}{{{3\pi ^2}}}\sum\limits_l {\left( {(2l + 1)\frac{1}{{{r^2}}}} \right.}  \times  \hfill \\
\;\quad \;\quad \left. {\int_0^{{k_{\rm F}}} {{k^2}} \left[ {\sin^2(kr - \frac{{l\pi }}{2} + {\delta _l}) - \sin^2(kr - \frac{{l\pi }}{2})} \right]} \right)dk \hfill \\
\end{gathered}.
\end{equation}
Performing integration along with applying some simplifications,
one gets
\begin{equation}\label{ind_ro_gup1}
\begin{gathered}
\delta {\rho ^{(1)}} =  -\frac{{ e}}{{3{\pi ^2}}}\sum\limits_l {\left( {(2l + 1)\left[ {\frac{{ - 2{k_{\rm F}}}}{{{r^8}}}\cos^2(2{k_{\rm F}}r - l\pi  + {\delta _l})} \right.} \right.}  \hfill \\
\,\,\,\,\,\,\,\,\,\,\left. {\left. {\, + \frac{{2{k_{\rm F}}}}{{{r^4}}}\sin(2{k_F}r - l\pi  + {\delta _l})\sin\;{\delta _l} + 2\cos\;{\delta _l}\;\sin\;{\delta _l}} \right]} \right) \hfill \\
\end{gathered}.
\end{equation}
It is clear from this equation that there are some additional
oscillating terms within the potential.
%%%%%%%%=====================
\section{conclusion}

In this paper, the Hartree-Fock model has been formulated
considering the effects of minimal length. The minimal length
modifies the minimal volume, and density of states in the
phase-space. Our results showed that the Hartree-Fock energy is
affected by the minimal length significantly, and the effect is
even considerable at lower momenta. In the Hartree-Fock model we
encounter two important problems; the Fermi velocity divergence,
and the bandwidth increases, depending on density. We presented an
efficient approach to overcome these deficiencies by considering
the minimal length effects. Our calculations showed that
considering the value of $\beta$ parameter as
$\mathcal{O}(r_s/a_0^2)$ with an appropriate constant, the
problems of the Hartree-Fock model will be controlled.

We also found out that minimal length (GUP) modifies the
Schr\"{o}dinger equation with some extra terms corresponding to
different orders of $\beta$ parameter and
$\sqrt{\beta}=\sqrt{\beta_0}l_{\rm pl}$. Therefore, in spite of
physical significance, we can modify the many-electron
Schr\"{o}dinger equation by considering $l_{\rm
pl}\sim\mathcal{O}(r_s/a_0)$ adjusted by appropriate values of
$\sqrt{\beta_0}$. Therefore, regarding GUP formalism and its
consequences particularly in the form of Eq. \eqref{schro_GUP}, we
can write the Schr\"{o}dinger equation up to the first order of
GUP corrections, as Eq. \eqref{hamiltonian_gup_1}.
%
%\begin{equation}
%\label{HF_corr1}
%\mathcal{H}=H_0+\sum_i{\frac{\beta}{m}p_i^4},
%\end{equation}
%
%where $H_0$ denotes the Hamiltonian of Eq.
%\eqref{Hamiltonian1}.

On the other hand, although the contribution of the relativistic
motion of electrons in the strong attractive field and near the
nucleus are small, they cannot be ignored \cite{REL}. Accordingly,
due to the order of magnitude of $v_{\bf F}$, we may expect to
have more terms in the expansion of kinetic energy, meaning that
the $\mathcal{O}(p^4)$ terms in Eq. \eqref{hamiltonian_gup_1} can
also be justified by the relativistic corrections. Therefore, a
many-body Hamiltonian like
$$H=H_0+\sum_{i}\alpha_ip_i^4,$$
where $\alpha_i$ are coefficients appeared in theory, is generally
acceptable. We obtained that, in our model, the HF problems can
mathematically be solved. Although, we considered the effects of
GUP on both the system Hamiltonian and its phase space, our
results indicate that the HF problems are solved, if only the
phase space changes are considered. Indeed, the phase space
modifications play a very impressive and dominant role below the
Fermi surface, and also in eliminating the HF problems compared to
that of the Hamiltonian correction.

Another approach to overcome the HF infinite velocity at Fermi surface
is to consider the screening potential and the Linhardth theory of
screening as well. In this approach, the effect of minimal length on the
screen potential has been also investigated in the light of
Lindhard theory. We showed that in the minimal length formulation,
one must modify the screening length and parameter as well as the
long range Friedel oscillations. Our investigation indicates that the
effects of minimal length on screening adds more oscillating terms
to the induced density.

%%%%%%%%%%%%%%%%%%%%%%%%%%%%%%%%%%%%%%%%%%%%%%%%%%%%%%%%%%%%%%%%%%%%%%%%%%%%%
\section{Acknowledgment}
The authors would like to appreciate the anonymous referees for
providing useful comments and suggestions that helped us to
improve the original version of our manuscript.
%%%%%%%%%%%%%%%%%%%%%%%%%%%%%%%%%%%%%%%%%%%%%%%%%%%%%%%%%%%%%%%%%%

\end{document}